\documentclass[structabstract]{aa}

\usepackage{natbib}                     
\usepackage{amsmath}                    
\usepackage{graphicx}                   
\usepackage[varg]{txfonts}              
\usepackage[english]{babel}             
\usepackage{nameref}                    
\usepackage[normalem]{ulem}             
\usepackage{paralist}                   

\def\ang{\phi}
\DeclareMathOperator{\atan}{atan}

\begin{document}

  \title{Detecting ring systems around exoplanets using high resolution spectroscopy: the case of \object{51\,Peg\,b}\thanks{Based on observations collected at ESO facilities
   under program 091.C-0271 (with the HARPS spectrograph at the ESO 3.6-m telescope, La Silla-Paranal Observatory).}}

  \author{N. C. Santos\inst{1,2}
          \and  J. H. C. Martins\inst{1,2,5}
          \and G. Bou\'e\inst{3}
          \and A. C. M. Correia\inst{3,4}
          \and M. Oshagh\inst{1}
          \and P. Figueira\inst{1}
          \and A. Santerne\inst{1}
          \and S. G. Sousa\inst{1}
          \and C. Melo\inst{5}
          \and M. Montalto\inst{1}
          \and I. Boisse\inst{6}
          \and D. Ehrenreich\inst{7}
          \and C. Lovis\inst{7}
          \and F. Pepe\inst{7}
          \and S. Udry\inst{7}
          \and A. Garcia Munoz\inst{8}
         }

  \institute{
          Instituto de Astrof\'isica e Ci\^encias do Espa\c{c}o, Universidade do Porto, CAUP, Rua das Estrelas, 4150-762 Porto, Portugal
          \and
          Departamento de F\'isica e Astronomia, Faculdade de Ci\^encias, Universidade do Porto, Rua do Campo Alegre, 4169-007 Porto, Portugal
        \and
        ASD, IMCCE-CNRS UMR8028, Observatoire de Paris, UPMC, 77 Av. Denfert-Rochereau, 75014, Paris, France
        \and
        CIDMA, Departamento de F\'isica, Universidade de Aveiro, Campus de Santiago, 3810-193 Aveiro, Portugal
        \and
        European Southern Observatory, Casilla 19001, Santiago, Chile
        \and
        Aix Marseille Universit\'e, CNRS, LAM (Laboratoire d'Astrophysique de Marseille) UMR 7326, 13388, Marseille, France
        \and
        Observatoire de Gen\`eve, Universit\'e de Gen\`eve, 51 ch. des Maillettes, CH-1290 Sauverny, Switzerland
        \and
        Scientific Support Office, Directorate of Science and Robotic Exploration, ESA/ESTEC, Keplerlaan 1, 2201 AZ, Noordwijk, The Netherlands
}

  \date{Received date / Accepted date }
  \abstract
  {}
  {In this paper we explore the possibility that the recently detected reflected light signal of 51\,Peg\,b could be
  caused by a ring system around the planet. } 
  {We use a simple model to compare the observed signal with the expected signal from a short-period giant planet with rings.
  We also use simple dynamical arguments to understand the possible geometry of such a system.}
  {We provide evidence that, to a good approximation, the observations are compatible with the signal expected from a ringed planet, 
  assuming that the rings are non-coplanar with the orbital plane. However, based on dynamical arguments, we also show that this configuration is unlikely. 
  In the case of coplanar rings we then demonstrate that the incident flux on the ring surface is about 2\% the value received
  by the planet, a value that renders the ring explanation unlikely.}
  {The results suggest that the signal observed cannot in principle be explained by a planet+ring system. We discuss, however, the possibility of using reflected light spectra to detect 
  and characterize the presence of rings around short-period planets. Finally, we show that ring systems could have already been detected by photometric transit campaigns, but
  their signal could have been easily misinterpreted by the expected light curve of an eclipsing binary.}
  \keywords{(Stars:) Planetary systems, Planets and satellites: detection, Techniques: spectroscopy, Stars: Individual: 51 Peg
}


  \maketitle
  
  -------------------------------------------------
  \section{Introduction}                                        \label{sec:Introduction}

The detection of the atmospheres of extrasolar planets is becoming one of the
major research topics in the exoplanet field \citep[for a recent review see][]{Burrows-2014}. Current technology and a detailed data analysis 
have already allowed the signature of the atmospheres of other worlds to be detected using different methods, such as transmission spectroscopy \citep[e.g.,][]{Charbonneau-2002,Vidal-Madjar-2003,Madhusudhan-2014}, occultations \citep[e.g.,][]{Deming-2005,Demory-2012}, and phase curve variations \citep[e.g.,][]{Angerhausen-2014}. These studies allowed several detailed analyses of exoplanet
atmospheres, including tracing of thermal or albedo maps of the planets \citep[e.g.,][]{Knutson-2007,Stevenson-2014,Demory-2013}.

Although a large majority of the exoplanet atmosphere studies involved space-based data, the 
use of ground-based instrumentation to detect exoplanet atmospheres is providing a growing amount
of information. This is particularly true concerning the use of high-resolution spectroscopic techniques.
Using both optical and the near-infrared (near-IR) wavelengths, these methods allowed  the spectrum to be probed in detail for 
several exoplanets \citep[for some examples see][]{Snellen-2010,Birkby-2013,Wyttenbach-2015}.

In a recent paper, \citet[][]{Martins-2015} have explored a new technique for detecting the signature of a high-resolution (optical) reflected light spectrum from
an exoplanet. This detection allowed estimation of the radius and albedo of the historical 51\,Peg\,b planet \citep[][]{Mayor-1995}, 
suggesting that this planet may be a high-albedo, inflated hot-Jupiter such as Kepler-7\,b \citep[with $A_g=0.35$][]{Demory-2013}.

The predicted star-to-planet flux ratio for a star$+$planet system be estimated from \citep[e.g.,][]{Seager-2010}:
    \begin{equation}
      \frac{F_{planet}}{F_{*}} = A_{g} \, g(\alpha) \left(\frac {R_{p}}{a} \right)^{2}
      \label{eq:FluxRatio}
    \end{equation}
where A$_g$ is the geometric albedo of the planet, $a$ the semi-major axis of the orbit, $g(\alpha)$ the
phase function, and R$_{p}$ the planetary radius. The FWHM 
(22.6$\pm$3.6\,km\,s$^{-1}$) and amplitude (6.0$\pm$0.4$\times$10$^{-5}$) of the detected planet-cross-correlation function (CCF), as detected in \citet[][]{Martins-2015}, when compared to the values of the stellar CCF (7.47\,km\,s$^{-1}$ and 0.48, respectively), 
would lead to a planet-to-star flux ratio of 3.8$\times$10$^{-4}$. By applying the equation above, we would then derive a geometric albedo 
far above unity if we assume a jovian-like radius for 51\,Peg\,b\footnote{Note that the area of a Gaussian 
is proportional to FWHM$\times$Amplitude}.

\citet[][]{Martins-2015} presented the results of some simulations suggesting that the observed (and larger than expected) 
FWHM broadening can be an artifact produced by non-Gaussian noise in the data, together with the fact 
that their detection was only possible at a three-sigma level. The authors thus only used a
comparison of the CCF depths to derive indicative values for the albedo and planetary
radius. (The two parameters are degenerate.) Martins et al. also suggested that the
parameters of the CCF should be taken just as indicative, even if  they consider the detection solid.

It is, however, interesting to explore the possibility that the observed CCF values
are real. In this case, what could explain such a wide and deep CCF as observed? One possibility for explaining the large 
FWHM would be the presence of strong winds or a very fast rotation velocity (close to the observed FWHM).
The signature of strong winds in exoplanets has indeed been observed using transmission spectroscopy
\citep[e.g.,][]{Snellen-2010}. However, such broadening of the CCF would also imply a decrease in its
depth. The wind explanation would thus not be able to explain the
total surface of the CCF. Some extra component would be needed. 
Besides this, though not discussed directly in the paper, the near-IR signal of 51\,Peg\,b detected 
by \citet[][]{Brogi-2013} using CO lines does not seem to show a clear sign of extra broadening (see their Figure\,3).
Winds should produce a broadening that is independent of the wavelength of the observations.

In the present paper we explore a new interpretation of the measurements done by Martins et al. In particular,
we try to understand if the signal detected could be explained by a reflective ring system around 51\,Peg\,b. 
The existence of such rings has already been demonstrated to be possible around hot-Jupiters \citep[e.g.,][]{Schlichting-2011}.
In Sect.\,\ref{sec:rings} we unveil our hypothesis based on the observations of \citet[][]{Martins-2015}, and
we explore how a ring system should be able to explain the observed signal at opposition, i.e., when the Earth, the star, and the planet are almost aligned {(in the same order)}. In this study, we assume that the rings are not coplanar with the planet orbit and provide the results as a function of the angle $\phi$ between the Earth-star-planet line and the ring plane.
In Sect.\,\ref{sec:tilt} we use dynamical constraints to investigate whether the necessary configuration corresponds to a physical scenario. 
A more detailed model of the reflected light from a coplanar ring system is then explored in Sect.\,\ref{sec:reflectivity}.
In Sect.\,\ref{sec:transit} we briefly show that rings may have already been detected using transit photometry,
though in some cases their signature could have been interpreted as caused by the eclipse of a 
stellar companion. We conclude in Sect.\,\ref{sec:conclusions}.

  \section{The case for rings: a simple model}                          \label{sec:rings}


In Eq.\,\ref{eq:FluxRatio} we denote the planet-to-star flux ratio expected for
a star$+$planet system. The presence of rings around the planet would alter this ratio, because these
would also reflect light toward the observer.

The orbital inclination of 51\,Peg\,b is close to 80 degrees
\citep{Brogi-2013,Martins-2015}. Moreover, in \citet{Martins-2015}, the
detection of the reflected light has been done almost at opposition.
Thus, to get a rough estimation of the reflected light in this condition, we assume a simple geometry of the problem in which the Earth, the star, and the planet are along a straight line. In this configuration,
 the light reflected by a 
ring system with inner radius $r_i$ and outer radius $r_o$ is, in a simple
approximation, given by the light reflected from a uniform inclined disk with radius
$r_o$ subtracted by the light reflected by a similar uniform (and inclined) disk with
a radius $r_i$. This reflected signal also depends on the geometric
albedo of the disk/ring system ($A_{g}^r$) and on the tilt angle $\phi$ between the Earth-star-planet line
and the plane of the rings at the moment of opposition. (The lower the value of $\phi$, the lower will be the ``cross section'' of the ring
as seen by the star {and by the observer.} By definition, we have $g(\alpha)=1$ at the maximum phase angle. As such, the total planet-to-star flux ratio of a planet with a (optically thick) 
ring system can be approximated by
    \begin{equation}
      \frac{F_{planet+ring}}{F_{*}} = A_{g} \left(\frac {R_{p}}{a} \right)^{2} + A_{g}^r g_r(\phi) \left[ \left(\frac {r_{o}}{a} \right)^{2} - \left(\frac {r_{i}}{a} \right)^{2} \right] \ ,
            \label{eq:FluxesRatio}
    \end{equation}
where $g_r(\phi) \approx \sin^2 \phi$ is a ``reflectivity'' function that depends on the tilt of the rings with respect to the line of sight (see section~\ref{sec:reflectivity}).
This model is very simplistic and only serves to understand whether a ring system can explain 
the observed signal.

To check that this configuration can explain the observed reflected light CCF of 51\,Peg\,b, 
the first thing we need is to constrain the possible values for $r_i$ and $r_o$ (i.e., of the inner and outer 
radii of a possible ring system around 51 Peg\,b).
Looking at the case of Saturn, we see that $r_{i}$ can be very close to the radius of the planet.
We see no reason for this to be different in the case of 51\,Peg\,b, so we thus consider that $r_{i}=R_{p}$.

From a dynamical stability point of view, to constrain the outer radius $r_{o}$ we use the Hill approximation. 
The Hill radius is derived from
    \begin{equation}
H = a\,\left(\frac{M_p}{3\,M_s}\right)^{(1/3)}
        \end{equation}
where $M_p$ and $M_s$ are the planet and stellar masses, respectively. Assuming that
the real mass of 51\,Peg\,b is 0.46\,$M_{Jup}$, that the mass
of its host, 51\,Peg, is 1.04 $M_{\sun}$ \citep[][]{Santos-2013}, and that the orbital separation 
is 0.052\,au \citep[][]{Martins-2015}, we then derive a value of H=5.9\,R$_{Jup}$. 
We should note, however, that several dynamical studies have pointed out that the outer edges
of the Hill sphere are unstable \citep[see discussion in][]{Schlichting-2011,Kenworthy-2015}.
If we assume that only regions within 2/3\,H are stable, then the outer edge of the ring
system around 51\,Peg\,b should be $\sim$4\,R$_{Jup}$.

We note, however, that for radii higher than the Roche radius, we should expect
that ring particles gather to form satellites. The Roche radius, 
below which a given satellite of density $\rho$
will break up,  can be derived from \citep[e.g.,][]{dePater-2010}
    \begin{equation}
 R_{roche} = 2.44\,R_{p}\,\left(\frac{\rho_p}{\rho}\right)^{1/3}
        \end{equation}
where $\rho_p$ is the density of the planet.  Assuming 
that 51 Peg\,b has a radius of 1.2\,R$_{Jup}$, we derive $\rho_p=0.6\,g\,cm^{-3}$. Considering $\rho=3\,g\,cm^{-3}$ 
(typical of rocks), this implies a Roche Radius of $\sim$1.5\,R$_{Jup}$. If we assume that 
beyond $2\,R_{roche}$ there should no longer be any ring particles\footnote{The rings of Saturn extend
beyond the Roche limit, in particular the E ring, though they are
essentially composed of micron and submicron particles \citep[][]{Hedman-2012}.}, this would imply $r_{out}\sim3.0\,R_{Jup}$,
a value lower than the one found above when assuming the Hill radius.

In addition, these values for $H$ and $R_{roche}$ also depend on the real mass for the planet. The values
above were computed assuming a mass of 0.46\,$M_{Jup}$ for 51\,Peg\,b. This corresponds to an orbital
inclination of 80\,degrees. However, the inclination found in
Martins et al. is affected by large error bars ($i=80^{+10}_{-19}$). For instance, for values of $i=61$\,degrees (the lower bound),
the real mass of 51\,Peg\,b would be 0.53\,$M_{Jup}$. This corresponds to a variation on the order of 
10\% in mass, a value that produces a minor effect in the derived $H$ and $R_{roche}$.

\begin{figure}[b]
\begin{center}
\includegraphics[width = \columnwidth]{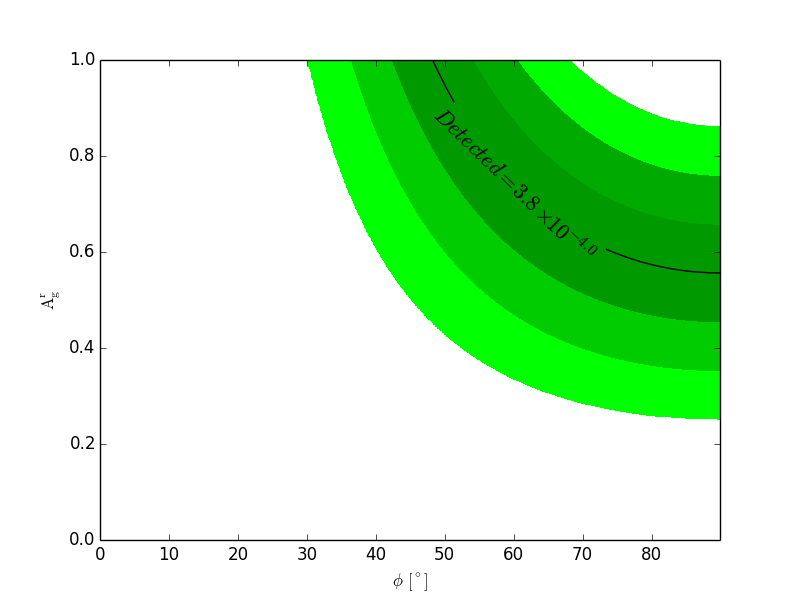}
\caption{Possible configurations of the inclination $\phi$ and A$_g^r$ given the detected signal-to-star flux ratio.
The color gradient denotes the 1-, 2-, and 3-sigma error bars.}
\label{fig:albedo}
\end{center}
\end{figure}

Adopting values for $r_o=3$ and $r_i=1\,R_{Jup}$, as derived above,
we then estimate a rough value for the expected $F_{planet+ring}/F_{*}$ from Eq.\,\ref{eq:FluxesRatio} (see Fig.\,\ref{fig:albedo}).
To do this, we also need to assume a value for A$_{g}$ and $A_g^r$, as
well as an inclination $\phi$ of the ring system with respect to the
Earth-star-planet line. We should note that $\phi$ is at most equal to the inclination
of the ring relative to the orbit and that this upper limit is only reached at
equinox.
Furthermore, this problem is highly degenerate. Different combinations of the
albedos and inclinations will be able to replicate the observed (planet + rings)-to-star flux ratio. 

To keep the different parameters within physically realistic values, we decided to fix $A_g$ to
0.3, a value that has been observed is several hot-Jupiters \citep[e.g.,][]{Cowan-2011,Demory-2013}.
As an example, using this value for $A_g$, $\phi=60$ degrees and a value of $A_g^r$=0.7, we 
can explain the observed flux ratio (assuming $r_o$ and $r_i$ of 3 and 1 R$_{Jup}$, respectively).

Increasing the inclination $\phi$ (i.e., increasing the angle between the
rings and the Earth-star-planet line) would imply that the projected area of rings would also increase, and 
lower values of $A_g^r$ would be necessary to explain the signal.  
In Fig.\,\ref{fig:albedo} we show the values of the inclination $\phi$ against A$_g^r$ that satisfy the observed flux ratio. 
In the figure, the errors on the detected signal-to-star flux ratio were computed from error propagation from 
the recovered values of the signal's amplitude and FWHM, i.e., $\frac{\Delta F}{F} = \sqrt{\left(\frac{\Delta Amp}{Amp}\right)^2 + \left(\frac{\Delta FWHM}{FWHM}\right)^2}$ where F , 
Amp, and FWHM are the signal-to-star flux ratio, the amplitude, and FWHM of the detected signal, respectively. The 
errors in the stellar parameters were ignored because they are much smaller than the ones of the detected signal.
Within the three-sigma error bars, possible solutions include pairs of $\phi$ and $A_g^r$ values as low as $\phi=40$ degrees and of the A$_g^r=0.6$.

It is not simple to understand what could be reasonable values for $A_g^r$. {Observations of Saturn's rings are not much help in this case, since they are rich in ices}. We found
no discussion in the literature about the expected albedo for silicates {and other refractory species} at the equilibrium temperature of
51\,Peg\,b, even if {such species} (e.g., SiO$_2$) are able to condense at the equilibrium temperature of 
51 Peg\,b \citep[$\sim$1200\,K -- see Fig.\,1 in][]{Schlichting-2011}. However, \citet[][]{Draine-1985} 
computed values for the single scattering albedo of silicates in the interstellar medium that are as 
high as 0.8 at optical wavelengths. These values could be increased if significant backscattering occurs near opposition,
as seen on Saturn's rings and other solar system bodies \citep[][]{Hameen-Anttila-1972,Buratti-1996,Verbiscer-2005}.

It is interesting to derive the expected ring rotational velocity and compare
it to the value of the observed FWHM. Assuming Keplerian rotation\footnote{$V_{rot}=\sqrt{\frac{G\,M_s}{r}}$, with $r$ the orbital radius 
of the ring particles and $G$ the gravitational constant, valid when the mass
of the rings is much less than the mass of 51 Peg\,b.}, the velocity of the rings should be 30\,km\,s$^{-1}$ at $r_i$ and 17\,km\,s$^{-1}$ at $r_o$. This shows that the presence of rings should also significantly broaden the CCF of the planet, as observed.
{
We should note, however, that the expected profile of the observed CCF may actually
present two different components: one produced by the light reflected on the planet disk 
(that will have a FWHM similar to the stellar CCF if the planet rotates slowly) and
the second produced by light reflected by the ring system, very likely with a broader CCF.
}

\citet[][]{Schlichting-2011} also call attention to the possibility that the Poynting-Robertson drag
slowly removes dust from a ring. 
{As discussed in \citet[][]{Burns-1984} and \citet[][and references therein]{Sfair-2009},
this effect implies that dust particles will remove orbital angular momentum and spiral
into the planet. For a typical hot-Jupiter and assuming a ring
orientation of 45 degrees with respect to the orbital plane, \citet[][]{Schlichting-2011} computed ring
lifetimes in the range of 10$^7$-10$^8$ years (see their Figure 4).} 
Assuming a lower
inclination and a higher density optically thick ring, however, this lifetime will increase strongly. 
We thus see no strong reason for a ring system around 51\,Peg\,b not to resist Poynting-Robertson drag. It is also worth noting that, as happens with Saturn, the 
presence of putative shepherd moons in the rings might greatly increase the lifetime of the 
rings; the presence of moons around hot-Jupiters is, however,
questionable \citep[][]{Weidner-2010}.

These numbers show that the observations reported in \citet[][]{Martins-2015} can in principle be
explained if we assume that 51\,Peg\,b has a ring system.
We should note, however, that these estimates should be seen mostly as qualitative. Our major goal at this stage was
to understand if the order of magnitude of the effect could be explained using the ring model.
For instance, it would be relevant to understand if such a scenario should have been detected by
existing {phase curve} observations using IR bands \citep[][]{Cowan-2007}.
The large uncertainties in the observed CCF parameters, the properties of a ring system (albedos, inclinations), 
and in the properties of the atmosphere of 51\,Peg\,b (e.g., the albedo and
wind velocities) precludes any deeper insight into this issue.

\section{Tilted rings?} \label{sec:tilt}

In the previous section we assumed that a putative ring system around 51\,Peg\,b could
have any tilt angle, following the suggestion of \citet[][]{Schlichting-2011}. As can be seen in Fig.\,\ref{fig:albedo}, this 
has a strong impact on our results. We therefore decided to verify  this assumption.

The initial spin state of the planet is unknown. 
The rotation period is supposed to be short, but the obliquity (the angle between the equator and the orbital plane, 
here denoted by $\varepsilon$) can assume any value, due to large impacts and planet-planet scattering at early stages 
in the formation process \citep[e.g.,][]{DonesTremaine-1993}.
However, due to the proximity of the star, the spin of hot-Jupiters slowly evolves until an equilibrium configuration is reached, 
corresponding to synchronous rotation and zero obliquity \citep[e.g.,][]{Hut-1980}.
The typical time scale $\tau$ for reaching this final equilibrium is given by \citet[e.g.,][]{Correia-2009}


\begin{equation}
\tau = \frac{P_{orb}}{9 \pi q} \frac{Q}{k_2} \ ,
\quad \mathrm{with} \quad
q =  \frac{M_s}{M_p} \left( \frac{R_p}{a}\right)^3 \ ,
\end{equation}

\noindent where $P_{orb}$ is the orbital period, $Q$  the dissipation quality factor, and $ k_2 $  the second Love number for the potential.
For Jupiter, astrometric observations provide $Q / k_2\approx10^5$ \citep{Lainey-2009}.
Adopting this same value for 51\,Peg\,b gives $\tau \sim10^5$~yr, strongly suggesting that it has reached its final configuration for the spin.
The same is true for all hot-Jupiters with $a < 0.1$~au. Values of $Q$ as high as $10^7$ have been proposed for stars \citep[][]{Penev-2012}, and this value could lead to $\tau\sim10^8$~yr if we assume $k_2\sim0.1$, as expected for giant planets \citep[][]{Yoder-1995}.


{
Ring systems are believed to have several possible origins: the result of impact events \citep[e.g.,][]{Tiscareno-2013},
captured objects or satellites that are tidally destroyed \citep[e.g.,][]{Charnoz-2009,Canup-2010},
or even remnants from planet formation \citep[though this last hypothesis is less likely --][]{Charnoz-2009b}.
}
In all cases, they settle in a special plane around the planet, called the Laplacian plane \citep[e.g.,][]{Lehebel-2015}.
It is usually defined as the plane normal to the axis about which the pole of a satellite's orbit precesses \citep{Laplace-1805}.
For circular orbits, the dynamics of the rings' particles is essentially governed by a single parameter, often called the Laplace radius \citep{Tremaine-2009}
\begin{equation}
R_L \equiv R_p \left( J_2 / q \right)^{1/5} \ .
\label{radlap}
\end{equation}
{For $r_o < R_L$, the rings can settle in the equatorial plane of the planet or in polar orbits. 
For $r_i > R_L$, the rings can only settle in the orbital plane of the system (implying $\ang=0$). }
For $r_i < R_L < r_o$, we expect a transition between the different regimes, called ``warped'' ring.


The parameter $J_2$ is related to the flattening of the planet.
For tidally evolved synchronous planets, we have \citep[e.g.,][]{Correia-2013}
\begin{equation}
J_2 = 5\,k_2\,q/6  \ .
\end{equation}
For Jupiter-like planets $k_2 \approx 1/2$ \citep[e.g.,][]{Yoder-1995}.
Using this value to compute the $J_2$ gives $R_L = 0.84 \, R_p$ for the Laplace radius (Eq.\,\ref{radlap}).
We thus conclude that for any hot-Jupiter $r_i > R_L$, so ring systems can only be observed in the orbital plane ($\ang=0$).

This result shows that ``warped'' rings are unlikely for close-in planets such as 51\,Peg\,b, 
except if we assume that the system is young and not yet synchronous. For a coplanar ring system, however, according
to the approximation presented in Eq.\,\ref{eq:FluxesRatio}, we expect no reflected light at all. The approximation
is thus no longer useful, though it hints that a ring configuration is likely not able to explain the reflected light
signal as observed in \citet[][]{Martins-2015}. It is, however, worth understanding what is the real amount of reflected light from
a ring system in such a situation.


\subsection{Reflectivity} \label{sec:reflectivity}

In this section, we no longer assume that the Earth is aligned
with the star-planet radius vector. It is then necessary to distinguish
two inclination angles of the rings' plane: the first one, denoted $\phi_i$, is computed with respect to the direction of
the star, while the second, $\phi_e$, is given relative to the line of sight. We note that at conjunction, $\phi_e = \phi_i = \phi$.
For a distant star (point source), the flux $F_r$ received by the rings depends only on the angle $\phi_i$ between the direction of the star and the rings' plane.
We thus have 
\begin{equation}
F_r(\phi_i) = F_p \sin \phi_i \ ,
\end{equation}
where $F_p$ is the flux that the ring would receive if it was
perpendicular to the incident light. The amount of flux reflected by the
ring in the direction of the observer also depends on the angle $\phi_e$
between the line of sight and the plane of the ring. If the scattering
is isotropic, the reflectivity reads (see Appendix)
\begin{equation}
g_r(\phi_i, \phi_e) = \frac{F_r(\phi_i)}{F_p} \sin \phi_e = \sin \phi_i
\sin \phi_e\ .
\end{equation}
 In the
approximation $\phi_i = \phi_e = \phi$, as in the previous section, we recover the dependency in $\sin^2 \phi$.
If a ring system around 51\,Peg\,b needs to be coplanar with the orbital plane, then $\sin^2 \phi = 0$, and the reflected light flux is thus only
due to the light reflected by the planet (Eq.\,\ref{eq:FluxesRatio} and Fig.\,\ref{fig:Fr_versus_phi}). 
However, for hot-Jupiters, the star cannot be seen as a point source, since the planet is close enough to receive light coming from the fraction of the stellar disk that illuminates the rings. It is thus interesting to estimate the total illumination, i.e., a more general expression for $F_r(\phi)$, and try to understand if this could actually be responsible for the observed signal.

\begin{figure*}[]
\begin{center}
\includegraphics[width = \textwidth]{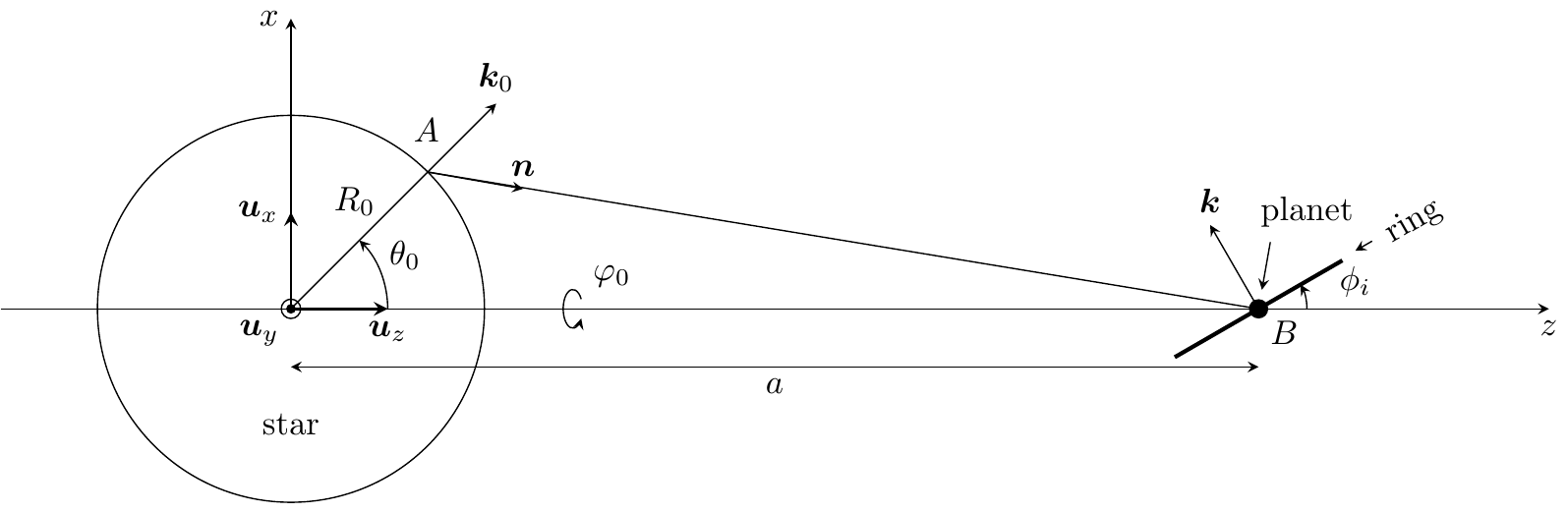}
\caption{Geometry of a system containing a hot-Jupiter with rings. $\vec
u_z$ is the unit vector along the radius vector of the planet relative
to the star, $\vec u_x$ is perpendicular to $\vec u_z$ in the plane
defined by $\vec u_z$ and $\vec k$ ($\vec u_x$ is not necessarily the
normal of the orbit). $\vec u_y$ complete the orthonormal base frame
$(\vec u_x, \vec u_y, \vec u_z)$.}
\label{rflux}
\end{center}
\end{figure*}

We let $I_s$ be the intensity emitted by the stellar surface and we assume that it is uniform, i.e., we neglect the limb-darkening. 
Then, the energy received by a ring element of surface $dA$ is given by 
\begin{equation}
dE_r (\ang_i) = I_s \iint
 (\vec k_0 \cdot \vec n) 
 \frac{(-\vec n\cdot \vec k)dA}{\|AB\|^2} 
 R_0^2 \sin\theta_0 \, d \theta_0\, d\varphi_0 \ , \label{renergy}
\end{equation}
where the double integral is computed over the portion of the stellar surface visible from a ring element. In this expression, $\vec k_0$ is the normal of the surface of the star at a given point $A$ of spherical coordinates $(R_0, \varphi_0, \theta_0)$, $B$ is a point of the ring, and $\vec n$ the unit vector along $AB$ (see Fig.\,\ref{rflux}).
For 51\,Peg\,b we have $a = 11\,R_0$, so we assume that $a \gg R_0$.
Moreover, as an order of magnitude, we only compute the incoming energy
at the center $B$ of the ring. The flux received by an element of the
ring is then defined as
\begin{equation}
F_r(\phi_i) \equiv dE_r(\phi_i) / dA\ ,
\label{Fr_dEdA}
\end{equation}
and $F_p$ is equal to $\pi I_s (R_0/a)^2$.
To compute the integral (\ref{renergy}), we consider two cases. If
the ring's inclination $\phi_i$ is greater than the angular radius of the
star $\phi_c = \atan(R_0/a)$, each element of the ring gets the light
from the full stellar disk: 
\begin{equation}
0\leq \varphi_0\leq 2\pi\ , \quad \mathrm{and} \quad
0\leq \theta_0 \leq \pi/2\ .
\end{equation}
In that case, at third order in $R_0/a$, the reflectivity of the ring
is still given by 
$ g_r(\phi_i,\phi_e) = \sin \phi_i \sin\phi_e$ (see Appendix).

\begin{figure}
\begin{center}
\includegraphics[width=\linewidth]{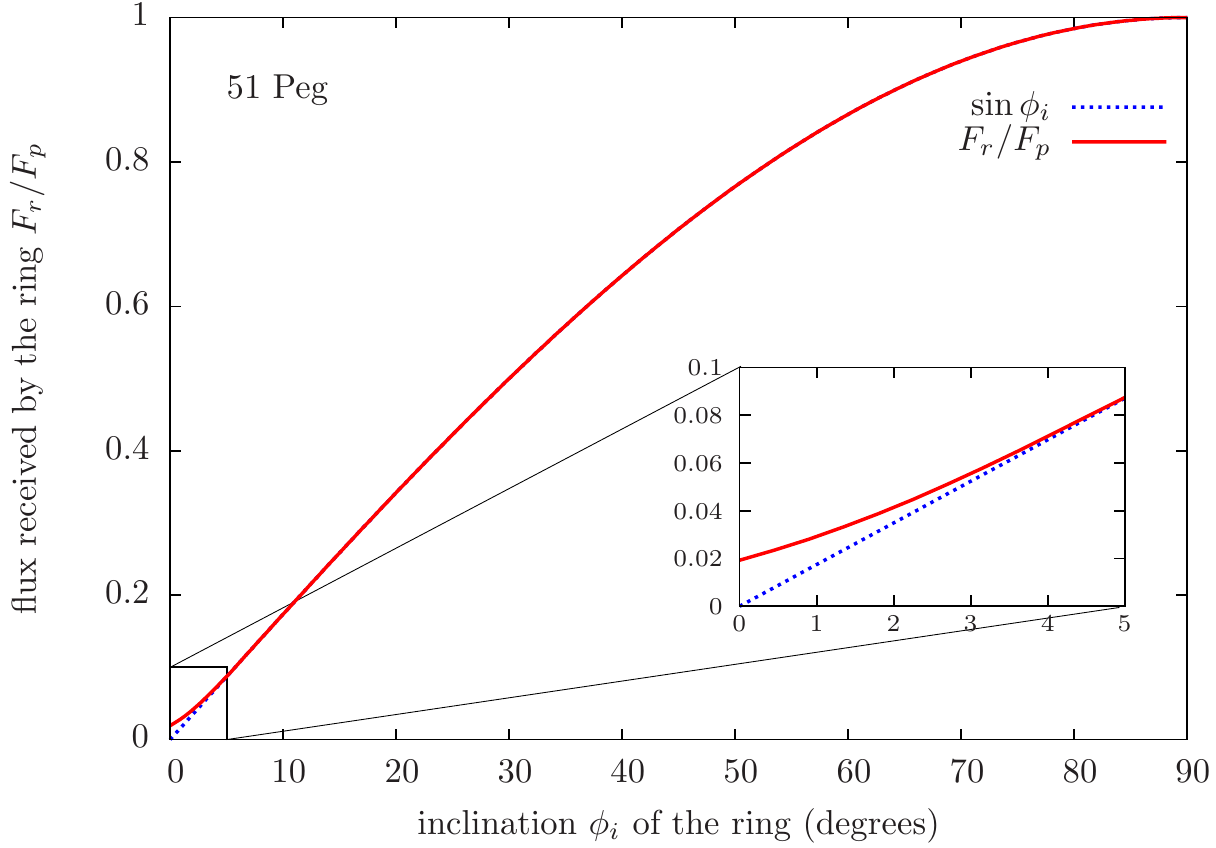}
\caption{Flux received by the ring as a function of the tilt angle.}
\label{fig:Fr_versus_phi}
\end{center}
\end{figure}

On the other hand, if the inclination $\phi_i$ of the ring is less than
$\phi_c$, a part of the stellar disk is occulted. In that case, the
reflectivity
becomes (see Appendix)
\begin{align}
g_r(\phi_i) &= \frac{\sin\phi_e}{\pi} \bigg\{ \left(\frac{\pi}{2}+\varphi_a\right)\sin\phi_i
+ \frac{2R_0}{3a}\cos\phi_i\cos^3\varphi_a \nonumber \\
& + \cos\varphi_a \sin\phi_i \bigg[
\sin\varphi_a - \frac{\pi}{2} \cos\varphi_a \sin\phi_i \nonumber \\
& + 
\frac{R_0}{a} \left(\pi \sin\varphi_a -
\frac{8}{3}\cos\varphi_a\sin\phi_i\right)
\cos\varphi_a\cos\phi_i
\bigg]\bigg\}\ ,
\end{align}
where $\varphi_a$ is defined as
\begin{equation}
R_0 \sin\varphi_a = a \tan\phi_i\ .
\label{eq.varphi_a}
\end{equation}
In particular, for small tilt angle $(\phi_i \ll R_0/a)$, we get
\begin{equation}
g_r(\phi_i,\phi_e) \approx \left(\frac{2R_0}{3\pi a} + \frac{\phi_i}{2}\right)\sin\phi_e\ .
\label{resflux}
\end{equation}

For 51\,Peg, in the limit of small tilt angles, $g_r \approx 0.02\sin\phi_e$, that is, the rings only receive about 2\% of the maximal flux computed at $\phi_i=90^\circ$.
This value is far too small to explain the observed signal as derived in Sect.\,\ref{sec:rings}.

\begin{figure}[t]
\begin{center}
\includegraphics[width = 9cm]{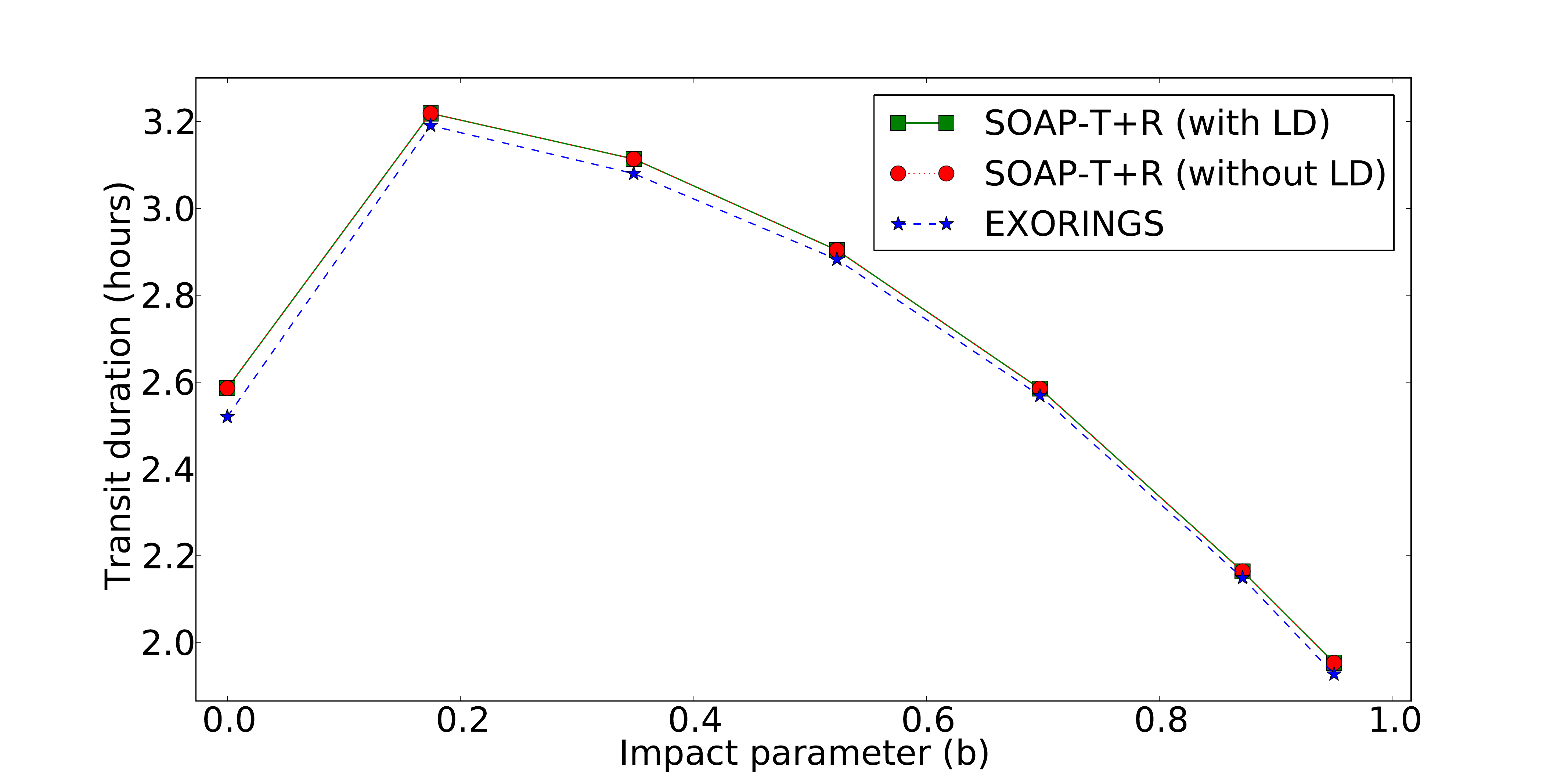}\\
\includegraphics[width = 9cm]{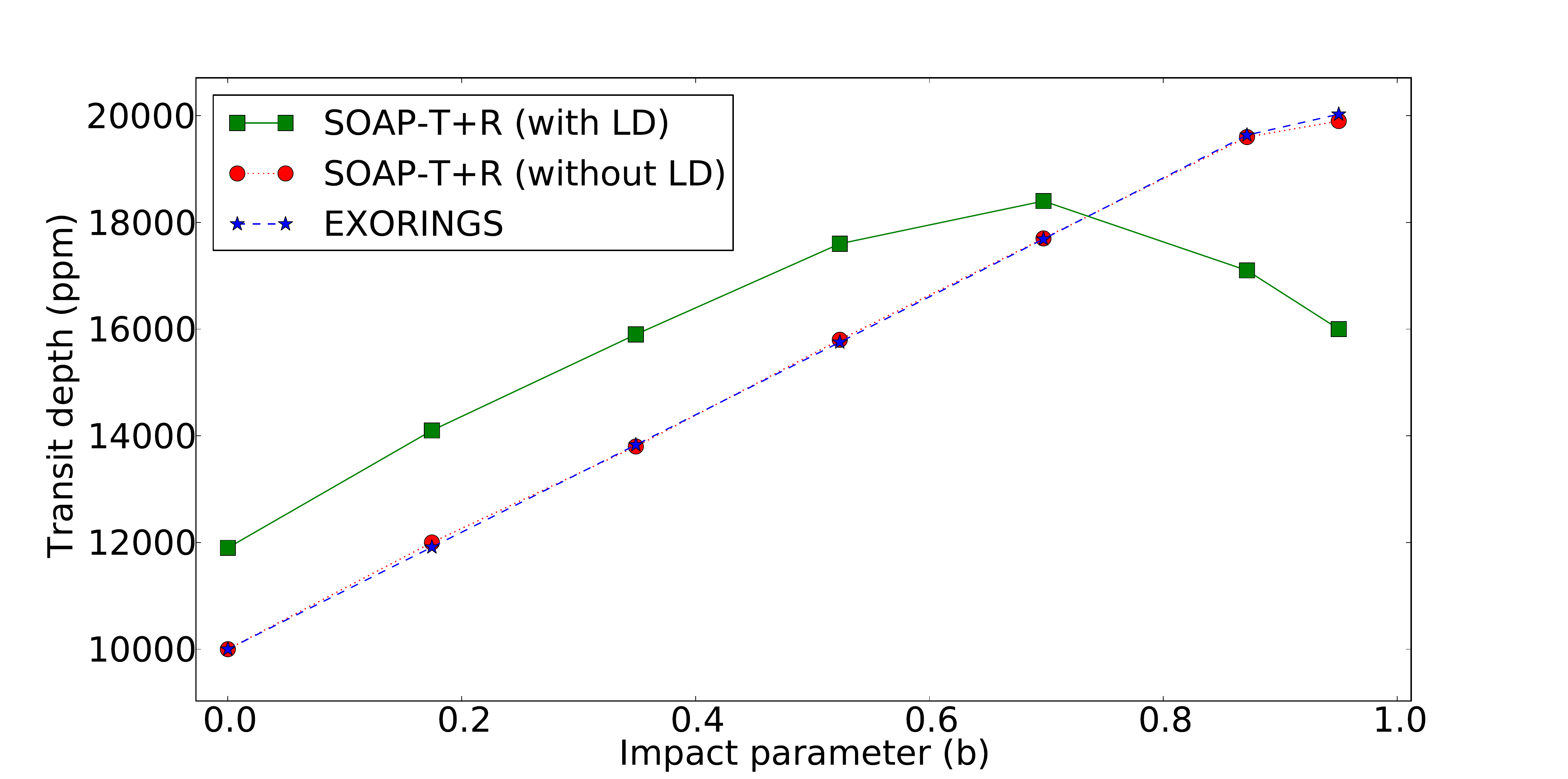}\\
\includegraphics[width = 9cm]{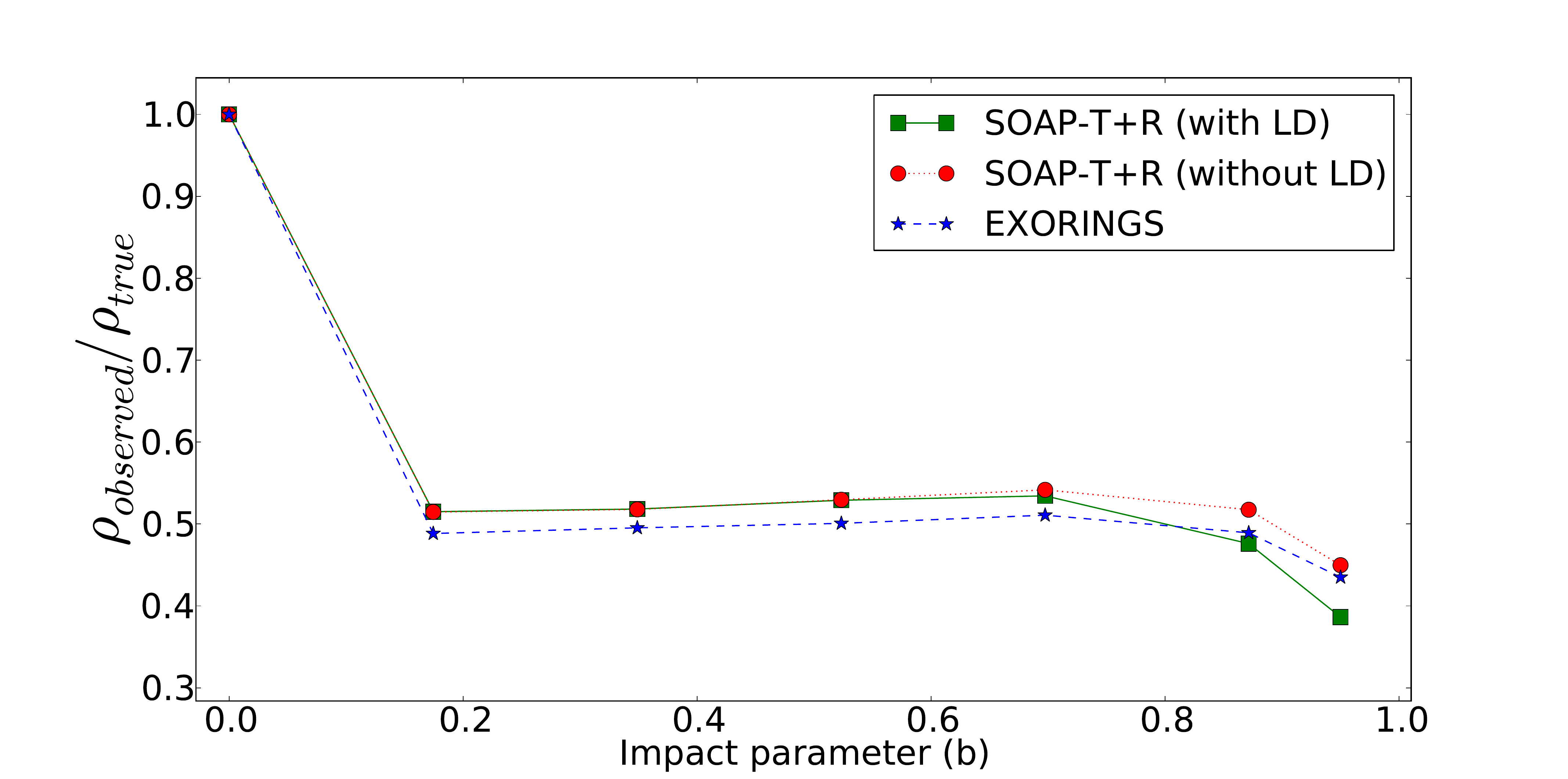}
\caption{Transit depth, duration signal, and derived stellar mean density for a transiting planet+ring system as a function of the impact parameter as computed using SOAP-T+R and EXORING.}
\label{fig:ring2}
\end{center}
\end{figure}

\begin{figure*}[t!]
\begin{center}
\includegraphics[width = 9.1cm]{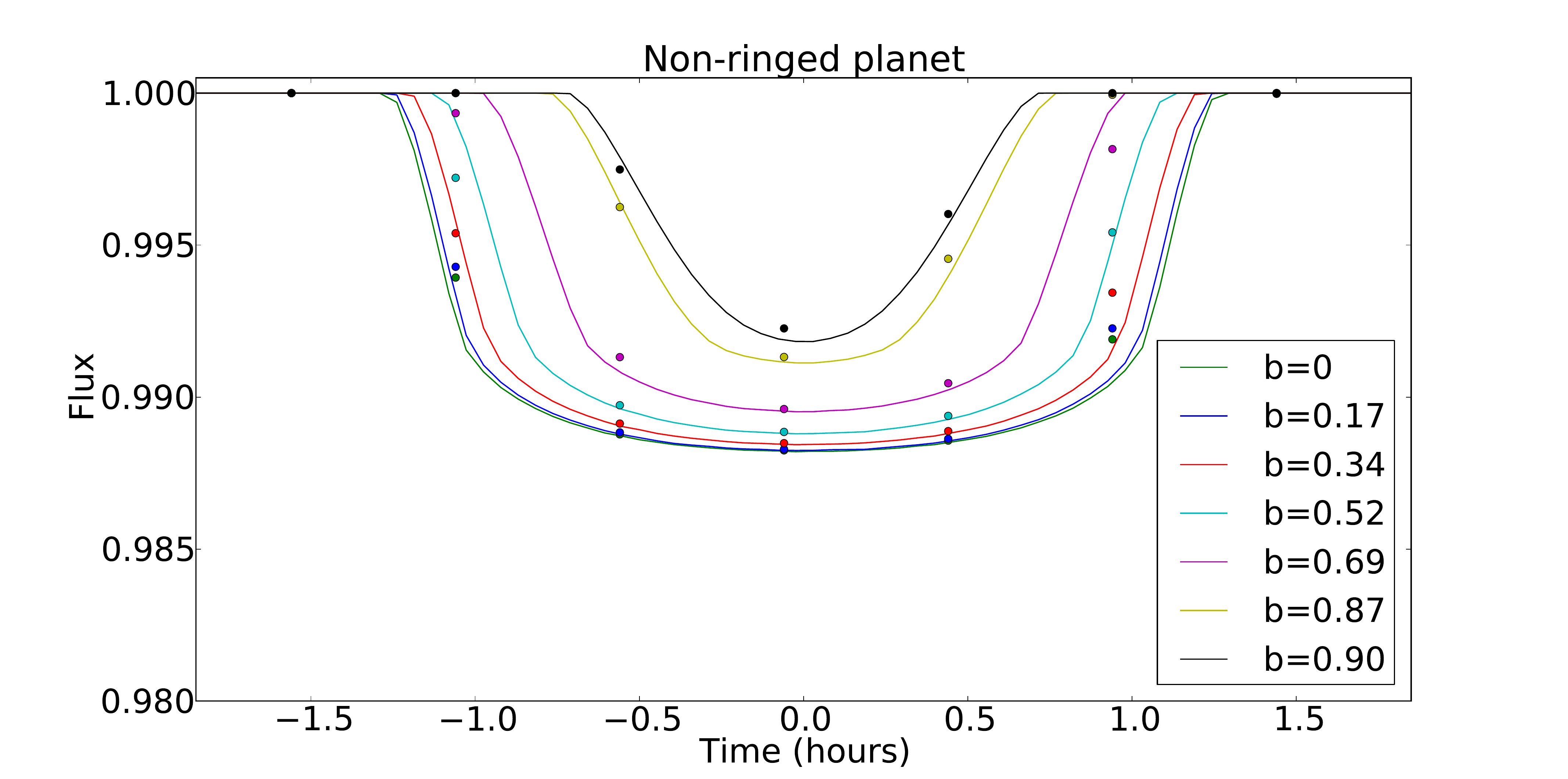}
\includegraphics[width = 9.1cm]{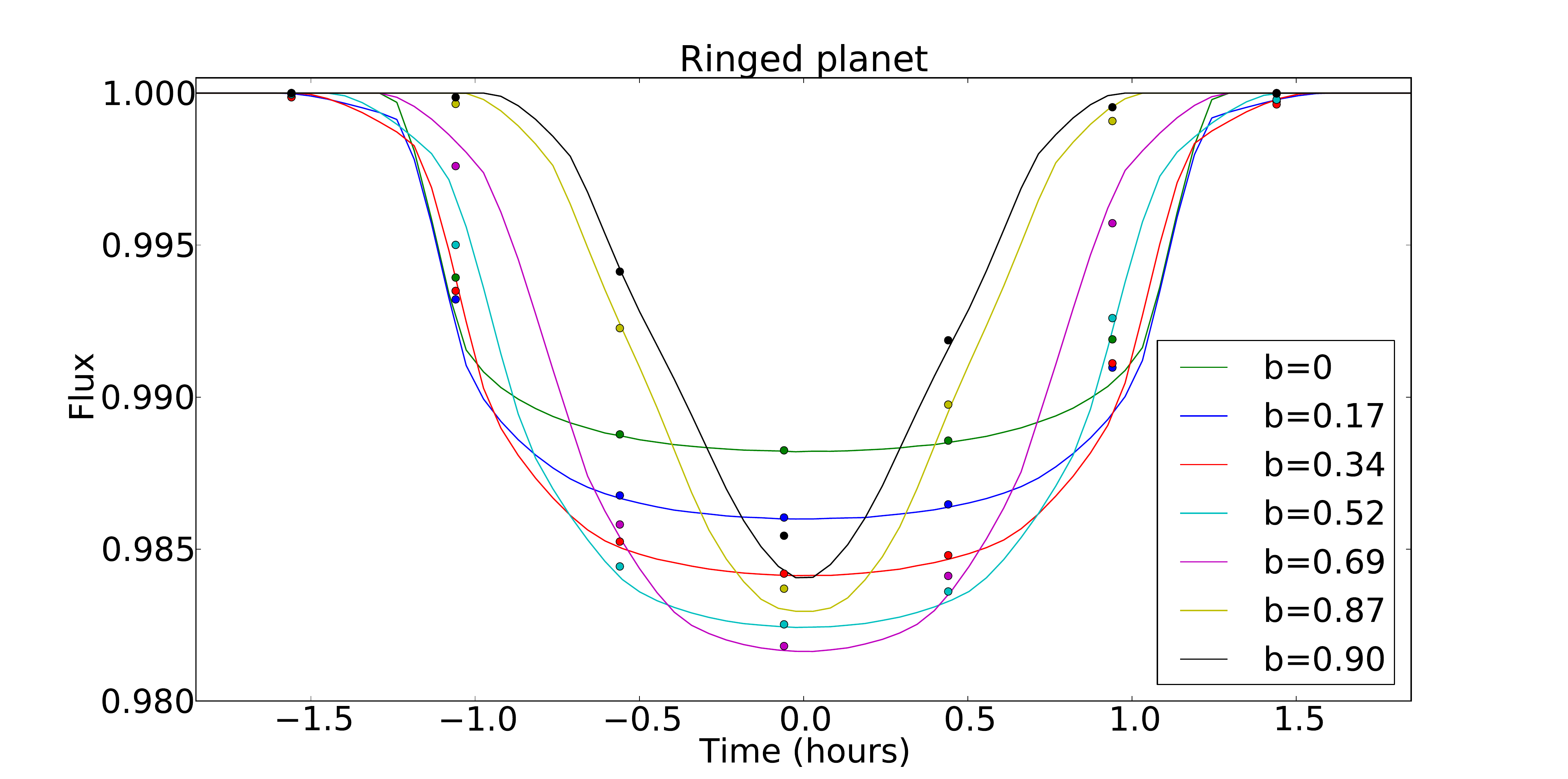}
\caption{Synthetic light curves simulating the light curves of transiting planets with (right) and without rings (left). The parameters of each planet are denoted in the inset. Dots correspond to data binned with a 30-minute cadence similar to long cadence of $Kepler$. Different system inclinations are used. In all the simulations, the ring is considered to be coplanar with the orbital plane. }
\label{fig:ring1}
\end{center}
\end{figure*}

%

  \section{Rings from transit surveys}\label{sec:transit}
  
Even though the results presented above do not support the ring hypothesis to explain the signal
observed in 51\,Peg\,b, the dynamical discussions presented also show that coplanar rings could be present 
around hot-Jupiters. One can thus wonder if rings are actually a frequent phenomenon around these sort of planets.

In a recent paper, \citet[][]{Kenworthy-2015} found evidence that the young pre-main-sequence star J1407 
(1SWASP J140747.93$-$394542.6 J1407) may have a planet with a massive ring system.
This case is not fully comparable to 51\,Peg\,b, in the sense that our target is much older and has a much shorter orbital period.
However, this example shows that present-day photometry is able to detect the presence of rings around exoplanets.

This issue has also also been discussed from a modeling point of view {\citep[][]{Arnold-2004,Barnes-2004,Dyudina-2005,Ohta-2009,Tusnski-2014}. }
In particular, simulations have shown that if massive rings are present in hot-Jupiters, the precision of transit surveys like $Kepler$ would 
have already allowed them to be detected. The question is then to understand if we have actually already detected rings photometrically 
but their existence passed unnoticed.

In a recent paper, \citet[][]{Zuluaga-2015} have shown that the presence of rings would produce (at least) two different effects.
One of these is that ringed planets would imply that
the value of the stellar surface gravity (or stellar density) derived from the light curve would be systematically smaller than
the one observed using asteroseismology, for example. 
Indeed, and although other parameters may be responsible for
the derivation of erroneous values of the stellar density from the transit light curve \citep[][]{Kipping-2014}, such a trend is 
expected if rings (even if not very massive/wide) are common around short-period exoplanets.

The other effect discussed in \citet[][]{Zuluaga-2015} is more ``obvious'': a ringed planet would trigger changes in the
transit light curve (with respect to a simple planet). In particular, transits should be deeper and longer. If the large rings are
present, the amplitude of the transiting signal could even be similar to the one expected 
for an eclipsing binary star.

To test these scenarios we modified the SOAP-T tool \citep[][]{Oshagh-2013} in order to add a planetary ring to the transiting planet (hereafter we call this code SOAP-T+R). SOAP-T was originally designed to generate the radial velocity variations and light curves for systems consisting of a rotating spotted star with a transiting planet. The model assumes that the rings are uniform and completely opaque and that they have an orientation with respect to the orbital plane. By comparing the transit light curves of SOAP-T+R with those of SOAP-T, we are able to recognize the impact of rings on the transit light curves. A full description of this code is beyond the scope of the present paper.

To check that the SOAP-T+R code was working properly, we compared its results with the ones obtained with the available EXORING code \citep[][]{Zuluaga-2015}. 
The obtained transit duration and depth 
are shown in Fig.\,\ref{fig:ring2} as simulated using both codes  as functions of the impact parameter for a ring system that is coplanar with the orbital plane. The comparison of the results obtained using SOAP-T+R and EXORING for the transit duration show very good agreement. On the other hand, the transit depth obtained from SOAP-T+R displays deeper transits than those using EXORING. This difference is explained by the fact that the stellar limb darkening is neglected in the EXORING
code, while in SOAP-T+R we consider a quadratic limb darkening coefficient close to the solar value. Indeed, if we
assume no limb darkening in SOAP-T+R, we obtain the same results as EXORING (see also Fig.\,\ref{fig:ring2}).

In Fig.\,\ref{fig:ring1} we show the results of our simulations after comparing the transit light curves produced by a planet (left) and by a planet with rings (right). Here we assume a ring system that is coplanar with the orbital plane and that has inner and outer radii of 1 and 4\,R$_{Jup}$, respectively. Quadratic limb darkening parameters u1=0.29 and u2=0.34 were used in this simulation (as expected for a Sun-like star). As seen in the figure, the impact of such a ring system can be quite significant. The most interesting result is that the shape of transit light curves of ringed planets (deep transit, long duration, and shape -- in most cases  ``V'' shaped)  look very much like the eclipse light curve of one eclipsing binary. Therefore, possible transiting planets with rings could have been observed by transit surveys, such as Kepler and CoRoT, however they could have easily been misclassified as false positive candidates.

{
We need to add, however, that the actual capability to distinguish between an eclipsing binary and a transiting ringed planet needs to be assessed 
by, for example, simulating the expected light curves in detail (including the different sources of noise) and investigating the residuals when 
fitting both models to the simulated data. We leave this detailed analysis to future studies.
}

The lower panel of Fig.\,\ref{fig:ring2} presents the impact of the ``unaccounted'' effect of a ringed planet on the derived stellar density as a function of planet impact parameter. To estimate this we used the stellar density as derived using the Eq.\,9 in \citet[][]{Seager-2003}. The plot shows that the stellar density is underestimated as we move toward higher impact parameter values. In this respect it is interesting to note that \citet[][]{Huber-2013} find that
the difference between the light-curve stellar density and the value derived using asteroseismology is 
a function of the impact parameter of the planet. We are not advocating, however, that rings are the
definite explanation for this trend. In any case, most of the systems in the Huber et al. paper are low-mass, small-radius,
and not Jovian in nature.



  \section{Conclusions}                                         
  \label{sec:conclusions}

In this paper we have explored the possibility that the reflected light spectrum observations of 51\,Peg\,b
\citep[][]{Martins-2015} can be explained if we assume that this hot-Jupiter has a ring system.
Using a simple model we showed that, overall, the observed signal can indeed be explained by a ring system under
the assumption that rings are tilted with respect to the orbital plane of the planet. We showed, however, that 
dynamical arguments suggest that in any synchronous hot-Jupiter like (we expect) 51\,Peg\,b, this configuration is unlikely. In the case of a ring 
system coplanar with the orbital plane of the planet, we also showed that the total amount of incident flux is about two orders of magnitude
smaller than the one needed to explain the observations.

The study shows, however, that the analysis of the reflected light spectrum from an exoplanet could
be a very interesting method of detecting rings around short-period systems, in particular those not transiting. 
{This approach can also complement the measurement of brightness variations along a phase curve, as already
proposed by \citet[][]{Dyudina-2005}}.
{Observations of apparently high-albedo planets that present broadened spectral lines or
line profiles showing two different components (see Sect.\,\ref{sec:rings}) could 
hint at the presence of rings around other planets.} A detailed model of the observed 
line-shapes could indeed provide relevant information about the system.

We also discussed the possibility that planets with rings could have been detected by space missions
like Kepler but simply discarded as binaries owing to the shape and depth of the transiting signal.
In this respect we propose that it would be interesting to obtain precise radial velocity measurements
of candidate binary stars from the Kepler field, in order to derive the masses of the companions.
The study of the light curves to search for transiting secondary binary like-signals with no signature of 
secondary eclipses or beaming and ellipsoidal effects \citep[][]{Mazeh-2012} 
could also help to select the best candidates.
We note that smaller ring systems could also be responsible for slightly deeper transits, 
leading to deriving an inflated radius for the transiting planet. A careful analysis
of the data could be relevant in such cases.

\begin{acknowledgements}
We would like to thank our referees, Dr. Sebastien Charnoz and the second anonymous referee, for the comments and suggestions that helped us to improve the quality of the paper. This work was supported by Funda\c{c}\~ao para a Ci\^encia e a Tecnologia (FCT) through the research grant UID/FIS/04434/2013. P.F., N.C.S., and S.G.S. also acknowledge the support from FCT through Investigador FCT contracts of reference IF/01037/2013, IF/00169/2012, and IF/00028/2014, respectively, and POPH/FSE (EC) by FEDER funding through the program ``Programa Operacional de Factores de Competitividade - COMPETE''. A.C. acknowledges support from CIDMA strategic project UID/MAT/04106/2013. PF further acknowledges support from Funda\c{c}\~ao para a Ci\^encia e a Tecnologia (FCT) in the form of an exploratory project of reference IF/01037/2013CP1191/CT0001.
A.S. is supported by the European Union under a Marie Curie Intra-European Fellowship for Career Development with reference FP7-PEOPLE-2013-IEF, number 627202. 
This work results within the collaboration of the COST Action TD 1308.
\end{acknowledgements}

\bibliographystyle{aa}
\bibliography{santos_bibliography}

\appendix

\section{Flux received by the ring}  \label{app:flux}

\begin{figure}
\begin{center}
\includegraphics[width=\linewidth]{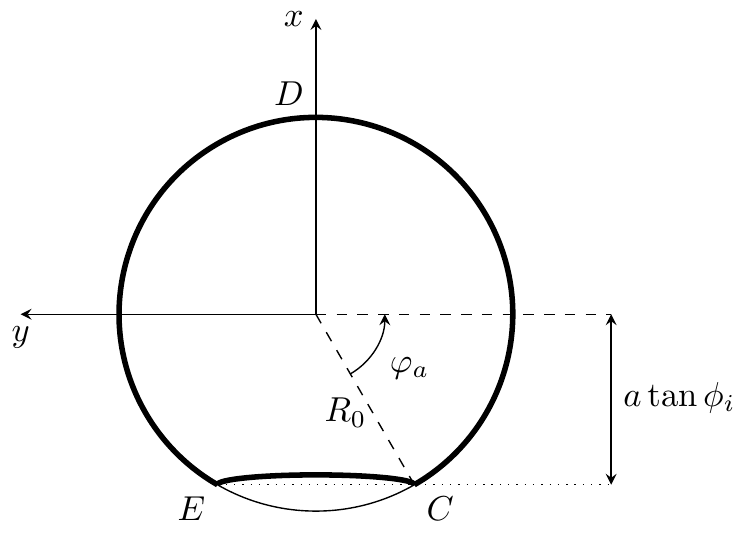}
\caption{Area enclosed by the thick curve $CDEC$ is the visible
part of the star seen from an element of the ring.}
\label{fig.visibleface}
\end{center}
\end{figure}

Here, we detail the computation of the flux $F_r(\phi_i)$ received by the
ring per unit area. The notation is the same as in Sect.~\ref{sec:reflectivity} of the main text (see also
Figure~\ref{rflux}). The general expression of the flux derived from
Eqs.~(\ref{renergy}) and~(\ref{Fr_dEdA}) is
\begin{equation}
F_r (\ang) = I_s \frac{R_0^2}{a^2} \iint
 \frac{a^2}{\|AB\|^2}
 (\vec k_0 \cdot \vec n) 
 (-\vec n\cdot \vec k)
 \sin\theta_0 \, d \theta_0\, d\varphi_0 \ ,
\end{equation}
where
\begin{equation}
\vec k = \begin{pmatrix}
\cos\phi_i \\ 0 \\ -\sin\phi_i
\end{pmatrix}\ , \quad
\vec k_0 = \begin{pmatrix}
\sin\theta_0 \cos\varphi_0 \\
\sin\theta_0 \sin\varphi_0 \\
\cos\theta_0
\end{pmatrix}\ , \quad
\vec n = \frac{a \vec u_z - R_0 \vec k_0}{\|a\vec u_z - R_0 \vec k_0\|}\
.
\end{equation}
The integrand in the expression of $F_r$ is of order unity. We expand it
at the first order in $R_0/a\ll 1$. We get
\begin{align}
F_r &= I_s \frac{R_0^2}{a^2} \iint
\bigg(
\sin\phi_i\cos\theta_0
+\frac{R_0}{a}\Big[
\sin\phi_i (3\cos^2\theta_0-1) \nonumber \\ &
+
\cos\phi_i\sin\theta_0\cos\theta_0\cos\varphi_0
\Big]
\bigg)\sin\theta_0 \,d\theta_0 \,d\varphi_0\ .
\label{eq.flux_integ}
\end{align}
We consider the most general case where each element of the ring only
sees a fraction of the stellar disk (see Figure~\ref{fig.visibleface}).
This case happens when the tilt angle $\phi_i$ is less than the angular
radius of the star $\phi_c = \atan(R_0/a)$. In this configuration, the
visible surface is delimited by two curves: the arc $CDE$ in the $xy$-plane of the
star and bounded by $-\pi/2-\varphi_a \leq \varphi_0 \leq
\pi/2+\varphi_a$, and the arc $EC$, which is half of a circle of radius
$R_0\cos\varphi_a$ in the plane of the ring.  For commodity, we recall
the definition of the angle $\varphi_a$ given in Eq.~(\ref{eq.varphi_a})
\begin{equation*}
R_0\sin\varphi_a = a \tan\phi_i\ .
\end{equation*}

To compute the surface integral (\ref{eq.flux_integ}), we make use of the
Stockes theorem that transforms a surface integral over $\Sigma$ into a
closed integral over its boundary $\partial \Sigma$ as
\begin{equation}
\iint_{\Sigma} \vec \nabla \times \vec A \cdot d\vec \Sigma = 
\oint_{\partial \Sigma} \vec A\cdot d\vec \ell\ .
\end{equation}
For this problem, we set
\begin{equation}
\vec A = \begin{pmatrix}
P \\ Q \\ R
\end{pmatrix}\ ,
\end{equation}
where
\begin{align}
P &= -\frac{1}{2} y \sin \phi_i + \frac{R_0}{a}(-yz\sin\phi_i-xy\cos\phi_i)\ , \nonumber \\
Q &=  \frac{1}{2} x \sin \phi_i + \frac{R_0}{a} xz\sin\phi_i\ , \\
R &=  0\ . \nonumber
\end{align}
For the line integral CDE, we use $\vec r = (x, y, z)$ with
\begin{equation}
x = \cos\varphi_0\ ,\quad
y = \sin\varphi_0\ ,\quad
z = 0
,\end{equation}
where $\varphi_0$ goes from $(-\pi/2-\varphi_a)$ to $(\pi/2+\varphi_a)$,. While for the line integral EC, we set $\vec r = (x, y, z)$ with
\begin{align}
& x = \cos\varphi_a\sin\phi_i\sin\psi-\sin\varphi_a\ , \nonumber \\
& y = \cos\varphi_a\cos\psi\ , \\
& z = \cos\varphi_a\cos\phi_i\sin\psi\ , \nonumber
\end{align}
where $\psi$ ranges from 0 to $\pi$. As a result, we get
\begin{align}
F_r(\phi_i) &= I_s \frac{R_0^2}{a^2} \bigg\{ \left(\frac{\pi}{2}+\varphi_a\right)\sin\phi_i
+ \frac{2R_0}{3a}\cos\phi_i\cos^3\varphi_a \nonumber \\
& + \cos\varphi_a \sin\phi_i \bigg[
\sin\varphi_a - \frac{\pi}{2} \cos\varphi_a \sin\phi_i \nonumber \\
& + 
\frac{R_0}{a} \left(\pi \sin\varphi_a -
\frac{8}{3}\cos\varphi_a\sin\phi_i\right)
\cos\varphi_a\cos\phi_i
\bigg]\bigg\}\ .
\label{eq.Fr_final}
\end{align}
In the case where $\phi_i > \phi_c$, Eq. (\ref{eq.Fr_final})
still holds if we set $\varphi_a = \pi/2$ so we get
\begin{equation*}
F_r(\phi_i) = \pi I_s \left(\frac{R_0}{a}\right)^2 \sin\phi_i\ ,
\end{equation*}
while for $\phi_i \ll \phi_c$, $\varphi_a \sim a\phi_i/R_0$ and
\begin{equation*}
F_r(\phi_i) \approx I_s \left(\frac{R_0}{a}\right)^2
\left(\frac{2R_0}{3a} + \frac{\pi}{2}\phi_i\right)\ .
\end{equation*}

\section{Reflectivity}
The reflectivity of the ring is computed by assuming an isotropic
scattering. Furthermore, it is assumed that given an incoming flux
$F_r$, only a fraction $A_g^r F_r$ is re-emitted in the visible
spectrum. Thus, the luminous intensity $I_r$ of the rings is uniform and
such that
\begin{equation}
A_g^r F_r(\phi_i) = \pi I_r\ .
\end{equation}
Besides this, the flux received on Earth from the disk is
\begin{equation}
F_\mathrm{ring} (\phi_i, \phi_e) = I_r
\frac{S_\mathrm{proj}(\phi_e)}{D^2}
\end{equation}
where the ratio of the projected surface $S_\mathrm{proj}$ of the rings
on the plane of the sky divided by the square of the distance $D$ to the
Earth represents the solid angle under which the rings are seen. As a
result, we get
\begin{equation}
F_\mathrm{ring} = \frac{A_g^r F_r(\phi_i)}{\pi}\times
\frac{\pi (r_o^2 - r_i^2)\sin\phi_e}{D^2}\ .
\label{eq.Fring}
\end{equation}
Moreover, the stellar flux $F_\star$ received on Earth is
\begin{equation}
F_\star = \pi I_s \left(\frac{R_0}{D}\right)^2 
        = F_p \left(\frac{a}{D}\right)^2\ ,
\label{eq.Fstar}
\end{equation}
where we used $F_p = \pi I_s (R_0/a)^2$. Combining Eq.~(\ref{eq.Fring})
and (\ref{eq.Fstar}), we get
\begin{equation}
\frac{F_\mathrm{ring}}{F_\star} = A_g^r \, g_r(\phi_i, \phi_e)
\left[ \left(\frac{r_o}{a}\right)^2 - \left(\frac{r_i}{a}\right)^2\right]\ ,
\end{equation}
with $g_r$ a function representing the reflectivity of the rings given
by
\begin{equation}
g_r(\phi_i, \phi_e) = \frac{F_r(\phi_i)}{F_p} \sin\phi_e\ .
\end{equation}

\end{document}